\documentclass[aps,prl,twocolumn,superscriptaddress,groupedaddress]{revtex4}  
\usepackage{graphicx}  
\usepackage{dcolumn}   
\usepackage{bm}        
\usepackage{amssymb}   
\usepackage{amsmath}   
\usepackage{mathtools} 
\usepackage{color}
\usepackage{ulem}

\usepackage[colorlinks,linkcolor=blue,urlcolor=blue,citecolor=blue]{hyperref}
 
 \usepackage{chemmacros} 

\begin{document}

\def\be{\begin{equation}}
\def\ee{\end{equation}}
\def\bea{\begin{eqnarray}}
\def\eea{\end{eqnarray}}
\def\l{\label}

\newcommand{\eref}[1]{Eq.~(\ref{#1})}%
\newcommand{\Eref}[1]{Equation~(\ref{#1})}%
\newcommand{\fref}[1]{Fig.~\ref{#1}} %
\newcommand{\Fref}[1]{Figure~\ref{#1}}%
\newcommand{\sref}[1]{Sec.~\ref{#1}}%
\newcommand{\Sref}[1]{Section~\ref{#1}}%
\newcommand{\aref}[1]{Appendix~\ref{#1}}%
\newcommand{\sgn}[1]{\mathrm{sgn}({#1})}%
\newcommand{\erfc}{\mathrm{erfc}}%
\newcommand{\erf}{\mathrm{erf}}%

\title{Thermodynamic uncertainty relation for first-passage times on Markov chains}

\author{Arnab Pal}
\thanks{Corresponding author}
\email{arnabpal@mail.tau.ac.il}
\affiliation{School of Chemistry, Raymond and Beverly Sackler Faculty of Exact Sciences, \& Center for the Physics and Chemistry of Living Systems. Tel Aviv University, 6997801, Tel Aviv, Israel}

\author{Shlomi Reuveni}
\email{shlomire@tauex.tau.ac.il}
\affiliation{School of Chemistry, Raymond and Beverly Sackler Faculty of Exact Sciences, \& Center for the Physics and Chemistry of Living Systems. Tel Aviv University, 6997801, Tel Aviv, Israel}

\author{Saar Rahav}
\email{rahavs@technion.ac.il}
\affiliation{Technion, Israel Institute of Technology, Haifa, Israel}

\date{\today}

\begin{abstract}
We derive a thermodynamic uncertainty relation (TUR) for first-passage times (FPTs) on continuous time Markov chains. The TUR utilizes the entropy production coming from bidirectional transitions, and the net flux coming from unidirectional transitions, to provide a lower bound on FPT fluctuations. As every bidirectional transition can also be seen as a pair of separate unidirectional ones, our approach typically yields an ensemble of TURs. The tightest bound on FPT fluctuations can then be obtained from this ensemble by a simple and physically motivated optimization procedure. The results presented herein are valid for arbitrary initial conditions, out-of-equilibrium dynamics, and are therefore well suited to describe the inherently irreversible first-passage event. They can thus be readily applied to a myriad of first-passage problems that arise across a wide range of disciplines.
\end{abstract}

\maketitle

\begin{figure}[t]
  \begin{center}
     \includegraphics[scale=0.6]{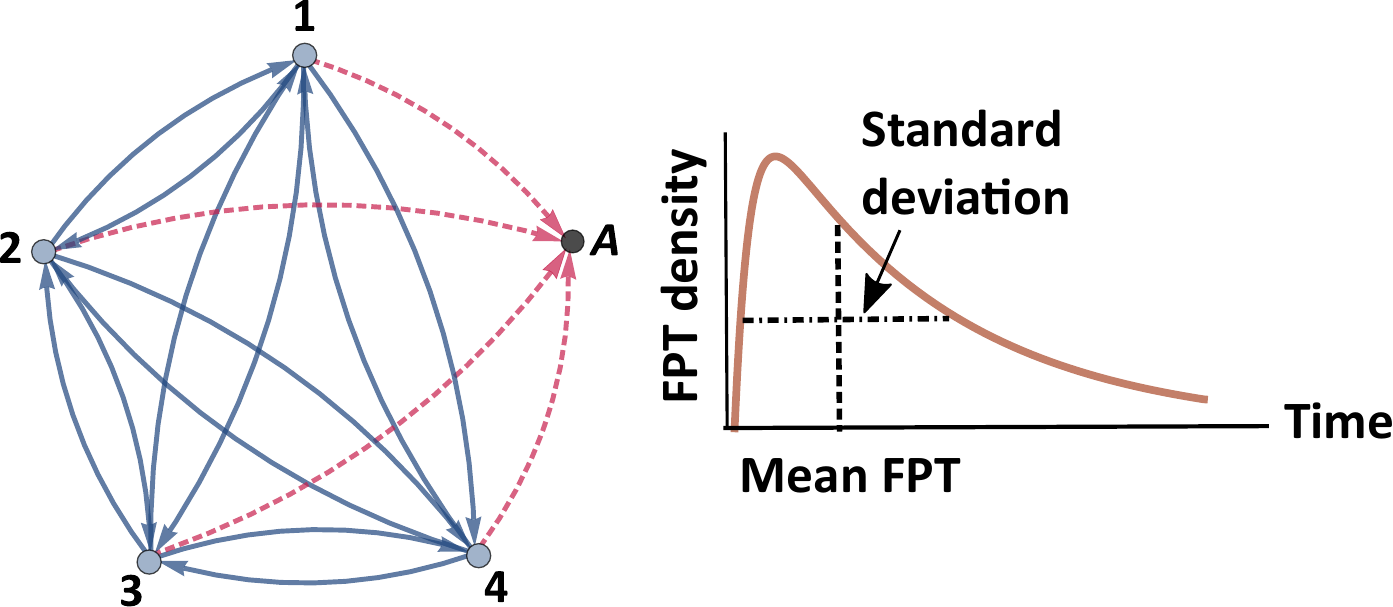}
    \caption{Schematic description of a Markov jump process on a network with an absorbing state  $A$. We develop a thermodynamic uncertainty relation that bounds relative fluctuations in the first passage time to $A$.}  \label{Fig:illustrate}
  \end{center}
\end{figure}

Many interesting phenomena occurring at the molecular scale can be viewed as first-passage processes \cite{redner2001guide}. From chemical reactions \cite{szabo1980first,szabo1984localized,Nitzan-book,Geometry-controlled,Gating} to enzyme catalysis  \cite{FP-EK,Ninio,FP-EK-2,Cao1,Enzyme1,Enzyme2,Enzyme3} and molecular search processes \cite{Mol_Search1,Mol_Search2,Mol_Search3,Mol_Search4,Mol_Search5} --- one is interested in the exact moment at which a distinguished event occurs: the first passage time (FPT) \cite{redner2001guide,Bray2013review,metzler2014first,klafter2011first}. Thermal agitations make FPTs random and much effort has been directed to determining their governing statistics in theory and experiment \cite{CV-expt,FPT-Theo-EXP1,FPT-Theo-EXP2,FPT-Theo-EXP3,FPT-Theo-EXP4,FPT-Theo-EXP5}. Of particular interest in that regard is the relative magnitude of FPT fluctuations around their mean value. This central gauge of randomness has found numerous uses and applications \cite{Gating,Enzyme1,Enzyme2,Enzyme3,Mol_Search5,FPT-Theo-EXP1,FPT-Theo-EXP2,FPT-Theo-EXP3,FPT-Theo-EXP4,Fluctuations1,Fluctuations2,Fluctuations3,Fluctuations4,Fluctuations5,Fluctuations6,Fluctuations7,Fluctuations8,Fluctuations9,Fluctuations10,Fluctuations11}. Yet, owing to conceptual and technical challenges, its relation with the underlying thermodynamics remains poorly  understood. 

The first passage event is often described as a completely irreversible step, and as a result such processes are inherently out-of-equilibrium. The theory of stochastic thermodynamics was developed to describe the nonequilibrium thermodynamics of small stochastic systems, such as molecular motors, enzymes, and nano-machines \cite{Sekimoto-book,ST-1,ST-2,ST-3,ST-4,ST-5,Pezzato2017}.  One of the celebrated results in the field is the thermodynamic uncertainty relation (TUR) \cite{TUR-1,TUR-2}. First conjectured by Barato and Seifert, the TUR states that the normalized fluctuations of currents are bounded from below by the inverse entropy production. The fundamental nature of the TUR has spurred an intensive research effort \cite{TUR-review,Seifert2019,Horowitz-2017,Pietzonka2017,Proesmans2017,Barato2018,Dechant2018,Hasegawa2019,Dechant2020,nonSS5,Uni,PRR-TUR,TUR-3,TUR-4}.

There has only been a handful attempts to extend the TUR to FPTs. Garrahan \cite{TUR-FP-2}, whose work was later on generalized by Gingrich and Horowitz \cite{TUR-FP-1}, derived a TUR for a fluctuating current that crosses a pre-determined threshold for the first time. More recently, Falasco and Esposito have derived a dissipation based speed limit on the mean arrival time to a target  \cite{MFPT-TUR}. Notably all the above-mentioned works focus on systems in steady state. In contrast, in first-passage processes, such as the one depicted in \fref{Fig:illustrate}, one is primarily interested in the transient dynamics. 
Moreover, the transition to the absorbing state is irreversible and therefore its entropy production is ill defined. These conceptual issues make an extension of the TUR to FPT processes both interesting and challenging. 

In this letter, we derive a TUR for the first-passage time to a site on a Markov chain. The result is derived from a more general set up that we have recently introduced to study  systems with unidirectional transitions \cite{PRR-TUR}. Like other TURs, the TUR derived herein bounds the relative fluctuations of the first-passage time, but its validity is not limited by the steady-state assumption and it can moreover be applied directly to systems with absorbing states. As this manuscript was being prepared Hiura and Sasa presented a kinetic TUR for first passage times \cite{TUR-FP-3} (see also \cite{KUR,Dechant-Sasa} for other kinetic TURs for currents). At the end of this letter, we show that the TUR derived in \cite{TUR-FP-3} follows from ours as a special case.

\textit{First passage on a continuous time Markov chain.---} On a Markov network, the probability to find the system in different states evolves according to the master equation
\begin{equation}
\frac{d \mathbb{P}}{dt}=\bm{\Gamma} \mathbb{P},
\label{ME}
\end{equation}
where $\bm{\Gamma}$ is the transition rates matrix. Its off diagonal
elements satisfy, $\Gamma_{ij}=K_{ij}$, where $K_{ij}$ is the rate of transitions from site $j$
to site $i$. The diagonal entries to $\bm{\Gamma}$  are given by 
$\Gamma_{jj}=-\sum_{i \ne j} K_{ij}$, namely the escape rates out of state $j$. For physical reasons one often assumes microreversibility, namely that if $K_{ij}>0$ then the reversed transition is also possible and $K_{ji}>0$. This principle is essential for the thermodynamical consistency of the model. However, when studying the kinetics of first passage processes one can not apply this principle to  transitions that lead to the absorbing state. Thus the challenge is to extend existing frameworks so as to take into account unidirectional transitions \cite{PRR-TUR}.

In what follows, we consider models with $N$ regular sites and one absorbing site labeled $N+1$. We also assume that all the transitions between the regular sites satisfy the principle of microreversibility. As a result, the transition rate matrix has the block structure
\bea
\bm{\Gamma}=
\begin{bmatrix}
\bar{\bm{\Gamma}} & \bm{0} \\
\bm{K}_\text{out} & 0
\end{bmatrix},
\eea
where $\bar{\bm{\Gamma}}$ accounts for the transitions between the regular states. $\bm{K}_\text{out} = (K_{N+1,1},K_{N+1,2},\cdots, K_{N+1,N})$ is a row vector containing the transition rates to the absorbing state.
For this class of models the distribution of first passage times is given by
\begin{equation}
    f_{T} (t) = \bm{K}_\text{out} e^{\bar{\bm{\Gamma}} t} \tilde{\mathbb{P}} (0),
\end{equation}
where $\tilde{\mathbb{P}} (0)$ is a vector containing the initial probabilities to reside in the $N$ regular sites. Once $ f_{T} (t) $ is known one can calculate and study its moments. In the following we will
focus on the mean and variance, and derive a bound on $CV^2= \text{Var}(T)/\left\langle T \right\rangle^2$, which is known as the coefficient of variation (CV) or randomness parameter of the first-passage time.

\textit{TUR for first passage time.---} A description of first passage that is in the spirit of stochastic thermodynamics can be
obtained by directly considering stochastic  realizations of the Markovian jump process  discussed above. In each such
realization the system starts at some initial site $i_0$, then jumps to
a site $i_1$ at time $t_1$, etc. An ensemble of all the realizations of duration $\mathcal{T}$
is obtained by assigning to each such realization its probability density. Now, let us define the functional
\begin{equation}
    T \left[\omega \right]= \sum_{i \ne A} \tau_i \left[ \omega \right],
    \label{eq:defT}
\end{equation}
where $\tau_i \left[ \omega \right]$ is the total time the realization $\omega$ spent in site $i$  throughout the observation time $\mathcal{T}$.
One notices that if the absorbing state is reached in a realization $\omega$, then 
$T\left[\omega \right]$ is simply the first passage time. Alternatively $ T \left[\omega \right]=\mathcal{T}$. The latter occurs with the survival probability $\mathcal{S} (\mathcal{T})$.
In finite and connected networks $\mathcal{S} (\mathcal{T})$ decays exponentially as $\mathcal{T} \rightarrow \infty$. In this limit the statistics of $T\left[\omega \right]$ is precisely  that of the first passage time.

A TUR for the first passage time can be obtained with the help of a more general inequality that we  recently derived for Markovian jump processes in systems with unidirectional transitions \cite{PRR-TUR}. For the functional (\ref{eq:defT}) it takes the form
    \begin{equation}
\text{Var}_\omega \left[ T(\omega)  \right] \geq \frac{\left[  \mathcal{T} \mathcal{S}(\mathcal{T})-\int_0^\mathcal{T}~dt~\mathcal{S} (t)  \right]^2}{\int_0^{\mathcal{T}} dt~\left[ \frac{1}{2}\sigma_{\text{rev}}(t)+J_{\text{uni}}(t) \right]},
\label{eq:TUR}
\end{equation}
where survival probability can also be understood as the ensemble averaged rate of change of the functional (\ref{eq:defT}): $\mathcal{S} (t)= \frac{d}{dt} \left\langle T[\omega] \right\rangle_t$. Notably, the bound in \cite{PRR-TUR} also holds for functionals that count transitions between states, and for models with more general transitions than the ones considered here.

To understand the denominator of Eq. (\ref{eq:defT}) we first note that transitions in our model can be divided into two sets. The set of unidirectional transitions, $E_1$, and the set of bidirectional transitions $E_2$. Importantly, every transition in the bidirectional set must have a reversed counterpart which is also in $E_2$. So if $(i\rightarrow j) \in E_2$ then also $(j\rightarrow i) \in E_2$.
There is no such restriction on transitions in $E_1$. In stochastic thermodynamics all the transitions
are often considered to be bidirectional due to microreversibility. One can then interpret
\begin{equation}
\sigma_{\text{rev}}(t)=\sum_{(j \rightarrow i) \in E_2} \frac{1}{2}\left[K_{ji} P_i(t)-K_{ij} P_j(t) \right] \ln \frac{K_{ji} P_i(t)}{K_{ij} P_j(t)},
\end{equation}
as the rate of entropy production from bidirectional transitions. In contrast, the contribution of
unidirectional transitions is 
\begin{equation}
J_{\text{uni}}(t)=\sum_{(j \rightarrow i) \in E_1}  {K}_{ij} P_{j} (t)~,
\end{equation}
which expresses their flux.

A crucial aspect of Eq. (\ref{eq:TUR}) is that it holds for arbitrary partitions of the model's
transitions into groups $E_1$ and $E_2$. For the first passage models we study here, the transitions
to the absorbing state must belong to the unidirectional set $E_1$ as they have no reverse counterpart. All other pairs of transitions that 
satisfy microreversibility can be placed in either $E_1$ or $E_2$. Thus, a reversible transition that would usually be placed in $E_2$ can be treated as a pair of irreversible transitions and put in $E_1$ instead. This freedom of choice gives rise to a set of bounds that emanate from Eq. (\ref{eq:TUR}). We emphasize that all these bounds are equally valid, and which one should be used is discussed at the end of this derivation.

To complete the derivation of a bound for the first passage time we consider the $\mathcal{T} \rightarrow \infty$ limit of Eq. (\ref{eq:TUR}). Using $\int_0^\mathcal{T}~dt~S(t)=\mathcal{T}S(\mathcal{T})+\int_0^\mathcal{T}~dt~t~f_T(t)$
and $\left\langle T \right\rangle = \lim_{\mathcal{T} \rightarrow \infty} \int_0^\mathcal{T}~dt~t~f_T(t)$,
we recast Eq. (\ref{eq:TUR}) as
\begin{equation}
   CV^2 \geq \frac{1}{\int_0^{\infty} dt~\left[ \frac{1}{2}\sigma_{\text{rev}}(t)+J_{\text{uni}}(t) \right]}.
    \label{eq:TUR-FP}
\end{equation}
Eq. (\ref{eq:TUR-FP}) is the central result of this paper. It shows that the coefficient of variation of the first passage time is bounded from below by an expression that combines the entropy production from the transitions in $E_2$ and the fluxes from the transitions in $E_1$. We now discuss different representations of the bound that arise from the freedom to place transitions in either of these groups.

\textit{Entropic bound.---} This form of the bound is obtained when all the bidirectional transitions are treated as such and put in group $E_2$. The remaining transitions are thus those that lead to the absorbing site $A$. Given that the system is initially distributed somewhere
among the sites $i \ne A$, and that $P_A (t \rightarrow \infty) =1$, one finds that 
\begin{equation}
    \int_0^\infty d t J_{\text{uni}} (t) = \int_0^\infty dP_A(t) = P_A (\infty) - P_A(0)=1.
\end{equation}
The entropy production from all other transitions is given by
\begin{align}
  \Sigma_\text{rev}=\frac{1}{2} \sum_{(j \rightarrow i) \in E_2} \int_0^\infty dt \left[ K_{ij} P_j (t) - K_{ji} P_i (t) \right] \ln \left[ \frac{K_{ij} P_j (t)}{K_{ji} P_i (t)}\right].
  \label{bi}
\end{align}
The entropic version of the bound is then given by 
\begin{equation}
    CV^2 \ge \frac{1}{\frac{1}{2} \Sigma_\text{rev}+1}.
    \label{eq:rev}
\end{equation}

\textit{Kinetic bound.---} The kinetic bound is obtained when all the reversible transitions, that were placed in $E_2$ in the entropic form of the bound, are treated as pairs of irreversible transitions and put in $E_1$.  As a result, the reversible entropy
production vanishes. In contrast, $\Sigma_\text{uni}$ is the integrated flux from all transitions, namely,
\begin{equation}
 \Sigma_\text{uni} = \int_0^\infty d t J_{\text{uni}} (t) = \sum_{(j \rightarrow i)} \int_0^\infty d t K_{ij} P_j (t).
 \label{uni}
\end{equation}
In this case one obtains the bound
\begin{equation}
    CV^2 \ge \frac{1}{\Sigma_\text{uni}}.
    \label{eq:uni}
\end{equation}

\textit{Mixed bounds.---}
Clearly Eqs. (\ref{eq:rev}) and (\ref{eq:uni}) follow extreme cases in which one considers the maximally possible number of transitions to be either bidirectional or unidirectional. Yet, one may also utilize the freedom to place some bidirectional transitions in $E_2$, while treating others as pairs of unidirectional directional transitions to be placed in $E_1$. This leads to a bound of the form
\begin{equation}
    CV^2 \ge \frac{1}{\frac{1}{2} \Sigma_\text{rev}+\Sigma_\text{uni}}.
    \label{eq:mix}
\end{equation}
Here $\Sigma_\text{rev}$ and $\Sigma_\text{uni}$ are similar to (\ref{bi}) and (\ref{uni}),
but with the summation restricted to the appropriate subset of transitions.

\begin{figure*}[t!]
\includegraphics[width=6cm]{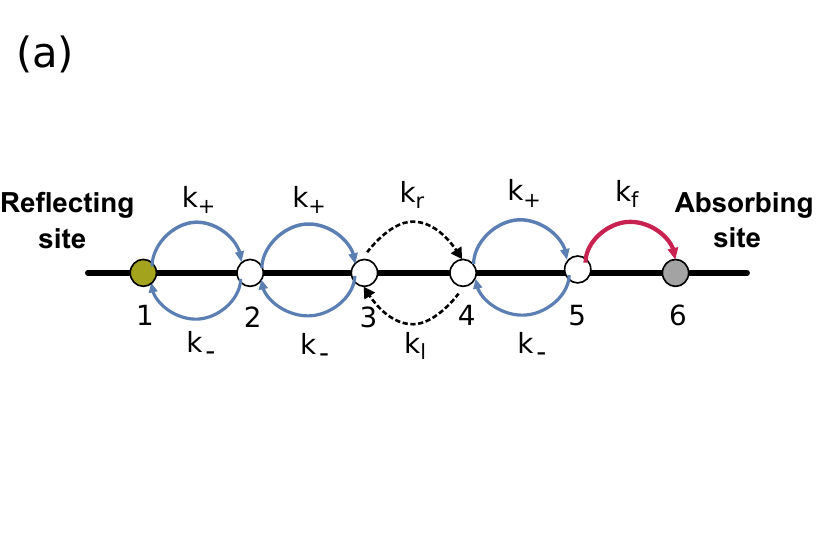}
 \includegraphics[width=5.75cm]{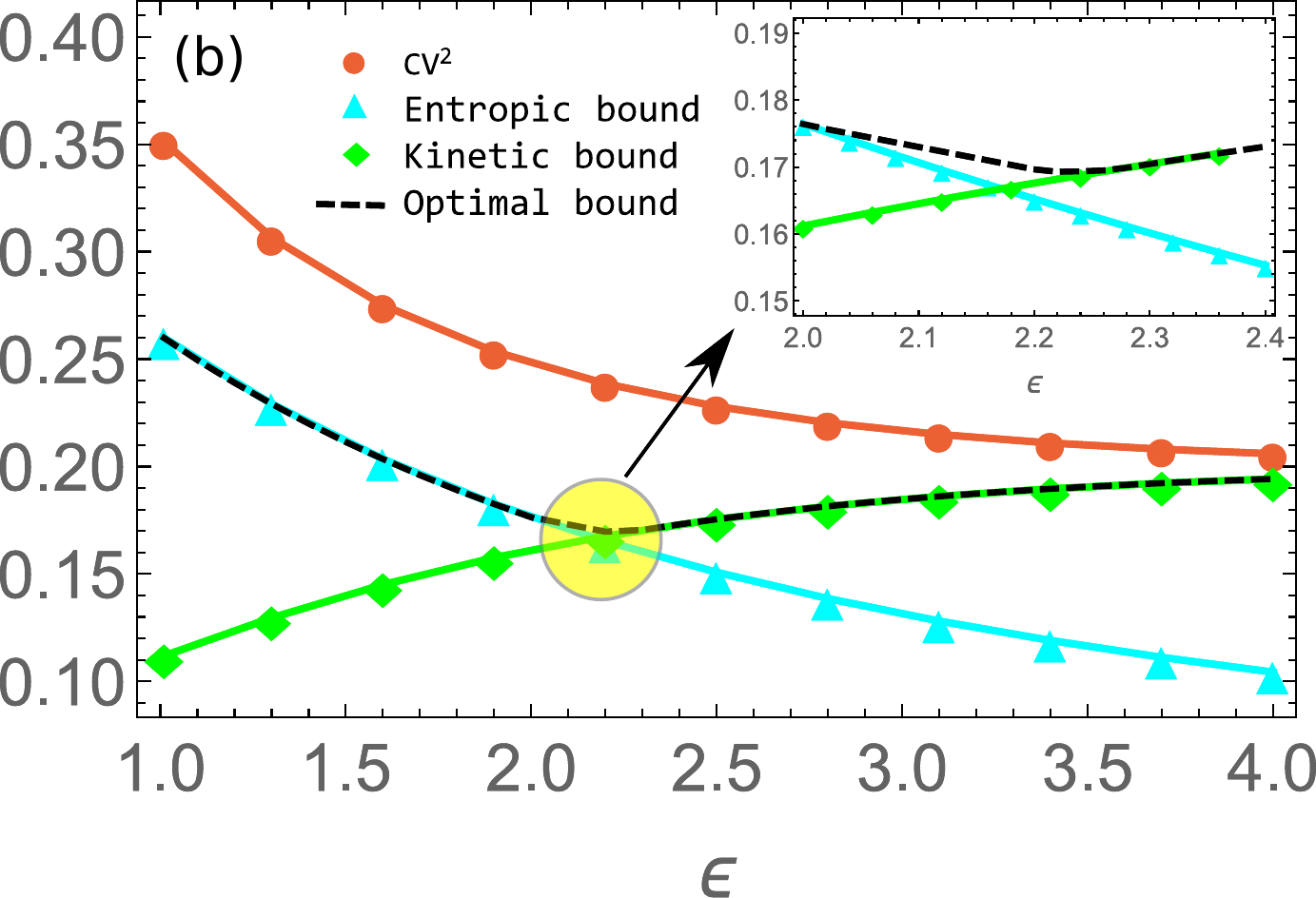}
 \includegraphics[width=5.65cm]{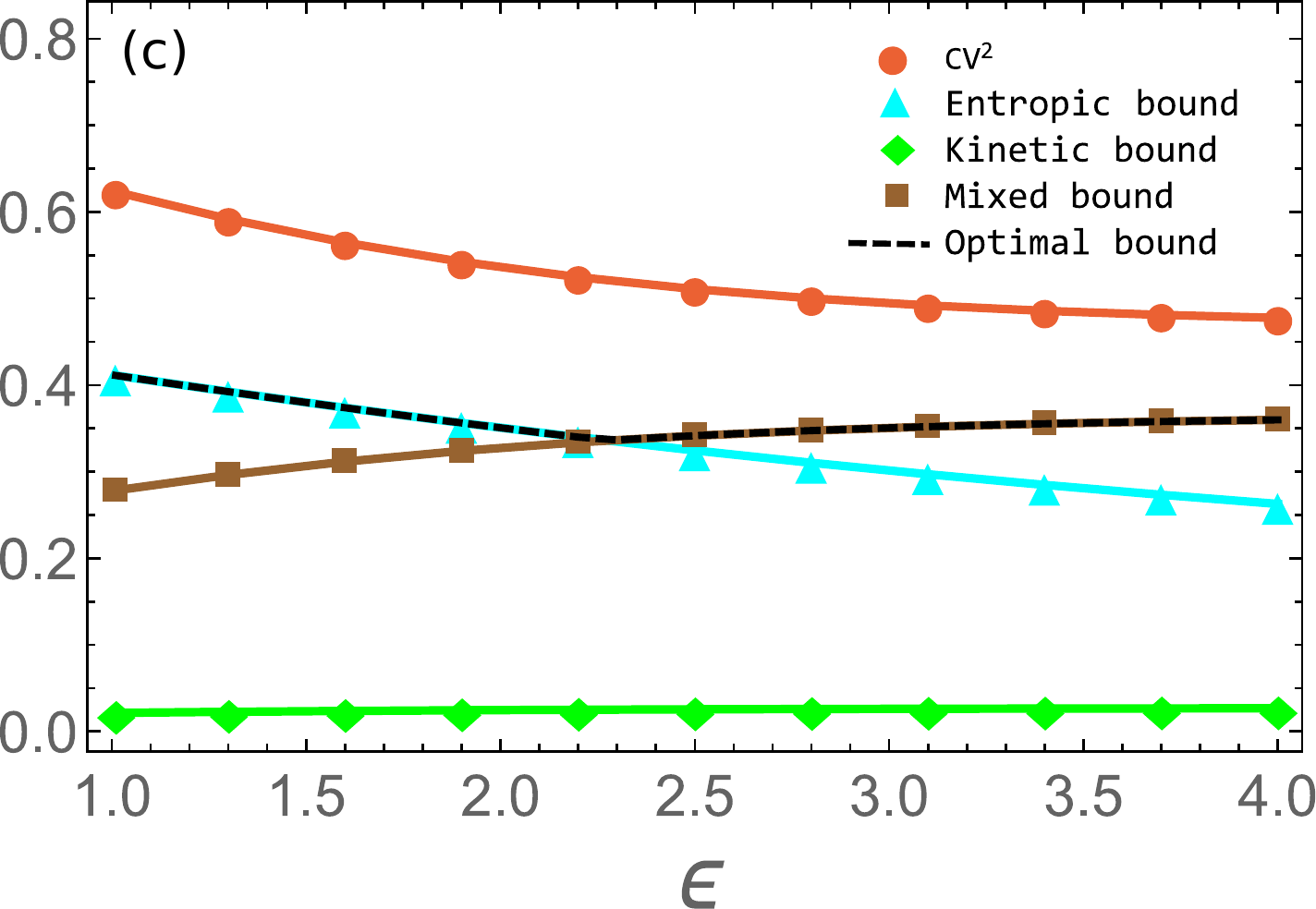}
 \caption{(a) A model of a  random walk on a finite 1d lattice. (b) The optimal bound on the CV of the FPT transitions from entropic to kinetic as motion on the lattice becomes heavily biased to the right (larger $\epsilon$, see main text for details). The inset focuses on the crossover region. The crossover from entropic to kinetic propagates from the edges of the system to its center. (c) The optimal bound on the CV of the FPT transitions from entropic to mixed, as the transition from the \{1,2,3\} block to the \{4,5\} block become more unidirectional (larger $\epsilon$, see main text for details).}
 \label{fig:num-sim}
\end{figure*}

\textit{The optimal bound.---}
The validity of several bounds raises a natural question: which bound should one use? We argue that the tightest bound is typically the most useful one. A simple algorithm allows to obtain the best bound from the family of bounds considered here. One has to scan over all pairs of sites $i,j$. For every pair of transitions $i \rightarrow j$ and $j \rightarrow i$ that can belong either to $E_1$
or to $E_2$ one calculates the quantities $\Sigma_1=\int_0^\infty dt \left[ K_{ij} P_j (t) +  K_{ji} P_i (t)\right]$ and $\Sigma_2=\int_0^\infty dt \left[ K_{ij} P_j (t) - K_{ji} P_i (t) \right] \ln \left[ \frac{K_{ij} P_j (t)}{K_{ji} P_i (t)}\right]$. If $\Sigma_1 < \Sigma_2$ one includes a contribution of
$\Sigma_1$ in $\Sigma_\text{uni}$. Alternatively, one includes a contribution of $\Sigma_2$ in $\Sigma_\text{rev}$.

We call this bound the \textit{optimal bound}, where the term optimal refers to the optimal partition of transitions into $E_1$ and $E_2$. Finding this optimal partition is relatively easy since each transition can be treated separately. Moreover, in many cases one can use physical intuition to figure out what to do with a given transition. For example, if a transition is expected to be almost detailed balanced, $K_{ij} P_j (t) \simeq K_{ji} P_i (t)$, then it is highly likely that $\Sigma_2 < \Sigma_1$ as illustrated below.

\textit{An illustrative example.---}
A simple model of a one dimensional random walker is used to elucidate our approach and to connect to  related results. The model consists of a linear chain of six sites, where the leftmost site is reflecting and the rightmost is absorbing [Fig. \ref{fig:num-sim}(a)]. The walker starts at the leftmost site, and jumps from site to site. All forward transitions in the model occur with rate $k_+$, except the $3 \rightarrow 4$ and $5 \rightarrow 6$ transitions that occur with rates $k_{r}$ and $k_f$ respectively. All backward transitions in the model occur with rate $k_{-}$, except the $4 \rightarrow 3$ transition that occurs with rate $k_{l}$. This parameterization of rates is flexible enough to exhibit several types of physically interesting behaviour. 

In \fref{fig:num-sim}(b), we demonstrate that the optimal bound transitions from the entropic form of \eref{eq:rev} to the kinetic form of \eref{eq:uni}. To do so, we set $k_+=k_r=k_f=1$, and  $k_-=k_l=e^{-\epsilon}$; and observe that all transitions are approximately unidirectional when $\epsilon$ is large. In this limit entropy production is very high. It is thus no surprise that replacing entropic with kinetic terms results in a tighter bound. On the other extreme, for small values of $\epsilon$, there is almost no bias which results in a quasi-equilibrium on the bulk of the lattice as probability gradually leaks to the absorbing site. In this limit, entropy production is much lower and the inclusion of entropic terms leads to tighter bound. (We use the term quasi-equilibrium loosely, as the flux to the absorbing site is not small.) For intermediate values of $\epsilon$, there is a crossover region where the behaviour is more complex leading to a mixed optimal bound that includes both entropic and kinetic terms (\fref{fig:num-sim}(b) inset).   

To show that a mixed bound can be optimal in large regions of the phase space, we make a different choice for the rates in \fref{fig:num-sim}(a): $k_+=k_-=1$, $k_r=k_f=1/5$, and $k_l=e^{-\epsilon}/5$. The lattice is then approximately divided into two blocks: that of sites \{$1,2,3$\} and that of sites \{$4,5$\}, with less frequent transitions between the blocks. When $\epsilon=0$, a lattice spanning quasi-equilibrium is formed and one expects the optimal bound to be entropic. In contrast, when $\epsilon$ is large the transition between lattice blocks is almost irreversible. In this case, a quasi-equilibrium is formed in each block separately, and one expects that the tightest bound will be achieved by taking entropy production terms for transitions within blocks and flux terms for transitions between blocks. \fref{fig:num-sim}(c) shows that this is indeed the case. For all parameters in this panel the kinetic bound performs poorly.

\textit{Summary and discussion.---}
We have derived a TUR for first passage times. Our approach allows to consider models with completely irreversible transitions which are commonly used to study first passage problems. Crucially, we derive mixed bounds that combine kinetic and thermodynamic contributions from different transitions on the Markov network at hand. While the manuscript was being prepared, a different but related approach was used in \cite{TUR-FP-3} to derive a TUR for first passage times. Their results also apply for models of the type studied here. However, note that the TUR derived there is equivalent to the fully kinetic version of our bound, and is thus a special case. The results presented in Fig. \ref{fig:num-sim}, as well as the considerations below, suggest that the ability to also consider mixed bounds is important.

Our results show an interesting connection to another well known bound
\begin{equation}
    CV^2 \ge \frac{1}{N},
    \label{eq:ASbound}
\end{equation}
where $N$ is again the number of kinetic states in the model (excluding the absorbing state). The bound (\ref{eq:ASbound}) was proven by Aldous and Shepp (AS) \cite{ASbound}, and is often used in the field of statistical kinetics \cite{FPT-Theo-EXP2,FPT-Theo-EXP3,FPT-Theo-EXP4}.
Barato and Seifert \cite{Barato2015} have shown that higher moments of first passage times also satisfy inequalities that depend on the number of states of the model.

We now discuss the relation between the TUR and the AS bound, with the help of the example depicted in Fig. \ref{fig:num-sim}. If all the transitions are irreversible, meaning $k_-=k_l=0$, then one is forced to use the kinetic bound. In this case $\Sigma_\text{uni}=N=5$
since the walker moves sequentially from the first site to the second, and continues moving right in this manner until it reaches the last site which is absorbing. We therefore recover the bound (\ref{eq:ASbound}).

Using the kinetic bound when $k_-$ or $k_l$ are non-vanishing will give $\Sigma_\text{uni}>N$, and thus a bound which is looser than the AS bound. Yet, a mixed TUR bound can significantly improve on the AS result. Consider, for example, a scenario with $k_l=0$ and $k_r,k_f \ll k_-,k_+$. These dynamics consist of fast local equilibration in two blocks of states: \{$1,2,3$\} and \{$4,5$\}, and rare transitions between the blocks. Here, a mixed bound, which uses the entropy production from the blocks and the fluxes of the unidirectional transitions, is tightest. In such a configuration, one expects $\Sigma_\text{uni}=2$ and $\Sigma_\text{rev} <1$: The mixed TUR behaves as if it is an AS bound of a coarse-grained model with two effective kinetic states. This is consistent with the results depicted in Fig. \ref{fig:num-sim}(c).

Importantly, Markovian models are coarse-grained approximations of underlying microscopic dynamics.
When several states of a model are almost always in local equilibrium with each other, thermodynamic consistency demands that we can coarse-grain them into one effective state, to obtain an even simpler model. The considerations above show how to get a TUR for coarse-grained models by appropriately partitioning transitions into two groups: uni- and bi-directional.

Finally, we note that several restrictions made on the model can be relaxed without many difficulties. The approach works for models with several physically distinct transitions between the same states. One simply needs to view these as separate transitions. It is also possible to study models with irreversible transitions inside the network of transitions (\cite{PRR-TUR,IFT-R-0,IFT-R,Work}). They do not need to lead to an absorbing state. Finally, one can also study models with several absorbing states. The derivation given in Ref. \cite{PRR-TUR} generically shows how to handle such models.

\textit{Acknowledgements.---}
Arnab Pal gratefully acknowledges support from the Raymond and Beverly Sackler Post-Doctoral Scholarship and the Ratner Center for Single Molecule Science at Tel-Aviv University. Shlomi Reuveni acknowledges support from the Azrieli Foundation, from the Raymond and Beverly Sackler Center for Computational Molecular and Materials Science at Tel Aviv University, and from the Israel Science Foundation (Grant No. 394/19).

\end{document}